\documentclass[a4paper,fleqn,usenatbib]{mnras}

\usepackage{newtxtext,newtxmath}

\usepackage[T1]{fontenc}
\usepackage{ae,aecompl}

\usepackage{graphicx}	
\usepackage{amsmath}	
\usepackage{amssymb}	

\title[Dark-matter-deficient galaxies in hydrodynamical simulations]{Dark-matter-deficient galaxies in hydrodynamical simulations}

\author[Jing et al.]{
Yingjie Jing,$^{1,2}$
Chunxiang Wang,$^{1,2}$
Ran Li,$^{1,2}$\thanks{E-mail: ranl@bao.ac.cn}
Shihong Liao,$^{1}$ 
Jie Wang,$^{1,2}$
\newauthor
Qi Guo,$^{1,2}$
Liang Gao$^{1,2,3}$
\\
\\
  $^{1}$Key Laboratory for Computational Astrophysics, National Astronomical Observatories, Chinese Academy of Sciences, Beijing, 100012, China\\
    $^{2}$School of Astronomy and Space Science, University of Chinese Academy of Sciences, Beijing, 100049, China\\
    $^{3}$Institute for Computational Cosmology, Department of Physics, University of Durham, South Road, Durham, DH1 3LE, UK\\
}

\date{Accepted XXX. Received YYY; in original form ZZZ}

\pubyear{2015}

\begin{document}
\label{firstpage}
\pagerange{\pageref{firstpage}--\pageref{lastpage}}
\maketitle

\begin{abstract}

Low mass galaxies are expected to be dark matter dominated even within their centrals. Recently two observations reported two dwarf galaxies in group environment with very little dark matter in their centrals. We explore the population and origins of dark-matter-deficient galaxies (DMDGs) in two state-of-the-art hydrodynamical simulations, the EAGLE and Illustris projects. For all satellite galaxies with $10^9<M_*<10^{10}$ M$_{\sun}$ in groups with $M_{200}>10^{13}$ M$_{\sun}$, we find that about $2.6\%$ of them in the EAGLE, and $1.5\%$ in the Illustris are DMDGs with dark matter fractions below $50\%$ inside two times half-stellar-mass radii. We demonstrate that DMDGs are highly tidal disrupted galaxies; and because dark matter has higher binding energy than stars,  mass loss of the dark matter is much more rapid than stars in DMDGs during tidal interactions. If DMDGs were confirmed in observations, they are expected in current galaxy formation models.
\bigskip

\end{abstract}



\section{Introduction}
In the standard $\Lambda$CDM framework, cold dark matter is the dominant mass budget of the Universe, but it can be less important at small scales. For example, for Milky Way-like galaxies and more massive ones, while dark matter dominates the total mass budget within the virial radius of their dark matter haloes, the central regions of these galaxies are dominated by baryons. Low-mass systems, for example  dwarf galaxies, however, have relatively shallower potential and suffer more from various physical processes, for instance, stellar feedback and UV ionization heating, and so can not make stars as efficient as high-mass systems. As a result, the central region of dwarf galaxies is predicted to be dark matter dominated. This expectation has been confirmed in observations of dark matter distributions of many nearby dwarf galaxies and also by modern hydrodynamical simulations \citep[e.g.][]{Schaller2015, Sawala2016}.

However, recent observations argue some of these low-mass galaxies may have very low dark matter fractions. For example, DDO 50, a galaxy in the catalog of LITTLE THINGS \citep{Oh2015}, has a stellar mass very close to the total dynamical mass, indicating a baryonic mass fraction 10 times higher than those of normal dwarf galaxies with the same maximum circular velocities \citep[also see][]{Oman2016}. In another recent observation by \citet{vanDokkum2018}, a low surface brightness galaxies, NGC1052-DF2, is reported to live in a system with only subdominant amount of dark matter. In their study, a dynamical analysis of globular clusters of this system shows that within a radius of 7.6 kpc, the total mass is $3.4\times10^8$ M$_{\sun}$ while the stellar mass is $2\times10^8$ M$_{\sun}$, implying that the dark matter mass is at least 400 times lower than galaxies of the same mass, but the results remain controversial \citep{Martin2018,Blakeslee2018,Wasserman2018,Laporte2019}.

In this short paper, we make use of two sets of state-of-the-art hydrodynamical simulations, the EAGLE \citep{Schaye2015,Crain2015} and Illustris \citep{Genel2014, Vogelsberger2014a,Vogelsberger2014b} projects, to explore whether or not the dark-matter-deficient galaxies (DMDGs) are allowed in current galaxy formation models. If yes, what is the physical origin of the formation of these DMDGs?

The structure of this paper is as follows. In Section \ref{sec:sim}, we describe briefly the simulation data used in this work. In Section \ref{sec:pop}, we show the fraction of DMDGs in our galaxy samples. We explore the formation history of DMDGs in Section \ref{sec:form}, and we summarize our results in Section \ref{sec:sum}. 

\section{simulations}
\label{sec:sim} 

The numerical simulations used in this study comprise two large hydrodynamical galaxy formation simulations, the Illustirs and EAGLE projects. Below we briefly describe these simulations. 

\subsection{Illustris simulations}

The Illustris project \citep{Genel2014,Vogelsberger2014a,Vogelsberger2014b,Sijacki2015,Nelson2015} consists of a series of cosmological hydrodynamical simulations performed with a moving-mesh code, \textsc{AREPO} \citep{2010MNRAS.401..791S}. The simulations assume a standard cosmological model with $\Omega_{\Lambda}=0.7274$, $\Omega_{\rm m}=0.2726$, $\Omega_{\rm b}=0.0456$, $\sigma_{8}=0.809$, $n_{\rm s}=0.963$, and $H_{\rm 0}=70.4$ km $\rm{s^{-1}}$ $\rm {Mpc^{-1}}$. In this study, we make use of the Illustris-1 simulation (hereafter Illustris), which follows the evolution of a cosmic volume of $(106.5 \rm{Mpc})^3$ from redshift $z=127$ to $z=0$. The mass resolutions are $m_{\rm DM}=6.26 \times 10^{6}{\rm M_{\sun}}$ for dark matter particles, and $m_{\rm b}=1.6 \times 10^{6}\rm{ M_{\sun}}$ for typical baryonic elements. The softening length for dark matter particles is fixed to $\epsilon_{\rm DM}=1.42$ kpc in comoving units, whereas the softening length for stars is limited to a maximum physical scale of $\epsilon_{\rm star}=0.71$ kpc. Haloes and subhaloes are identified with the friends-of-friend \citep[FOF,][]{Davis1985} and SUBFIND \citep{Springel2001} algorithm respectively. The merger trees of subhaloes are constructed by the SUBLINK algorithm; see \citet{Rodriguez-Gomez2015} for details.

\subsection{EAGLE simulations}

The EAGLE project \citep{Crain2015,Schaye2015} comprises of a suite of hydrodynamical simulations performed with an $N$-body Tree-PM smoothed particle hydrodynamics code, \textsc{GADGET-3} \citep{Springel2005}. The data of the simulation have been publicly released; see \citet{McAlpine2016} for details. In this work, we use the reference model run Ref-L0100N1504 (hereafter EAGLE) which follows the evolution of a volume of $(100 \mathrm{Mpc})^3$. The particle masses for initial gas and dark matter particles are $m_{\rm b}=1.8\times 10^{6}$ M$_{\sun}$ and $m_{\rm DM}=9.7\times 10^{6}$ M$_{\sun}$, respectively. The simulation adopts a comoving softening length of 2.66 kpc at early time, and switches to a fixed value of 0.7 kpc in physical scale after $z=2.8$ for dark matter and baryonic particles.

We also use the Ref-L0025N0752 run (hereafter EAGLE-highres) which has a higher mass resolution of $m_{\rm b}=2.26\times 10^{5}$ M$_{\sun}$, $m_{\rm DM}=1.21\times 10^{6}$ M$_{\sun}$ but a smaller volume of $(25 \mathrm{Mpc})^3$. The softening length for EAGLE-highres is 
$1.33$ comoving kpc initially, and is fixed to 0.35 physical kpc after $z=2.8$. All runs adopt a flat $\Lambda$CDM cosmology with parameters given by Planck results, i.e. $\Omega_{\rm{m}} = 0.307$, $\Omega_{\Lambda} = 0.693$, $\Omega_{\rm{b}} = 0.04825$, $n_{\rm s} = 0.9611$,  $\sigma_{\rm{8}} =0.8288$, and $H_{0} = 67.77$ km s$^{-1}$ Mpc$^{-1}$ \citep{Planck2014p1, Planck2014p16}. Similar to the Illustris simulations, haloes and subhaloes in the EAGLE simulations are identified by FOF and SUBFIND algorithms. The merger trees of subhaloes are constructed by the D-Trees algorithm \citep{Jiang2014}.

\section{dark matter fraction of satellite galaxies}
\label{sec:pop}
As the reported dark-matter-deficient dwarfs are discovered in group of galaxies, we select satellite galaxies in galaxy group environments to see whether or not they exist in modern cosmological hydrodynamical simulations. More specifically, we select satellite galaxies (belonged to FOF) with stellar mass $10^9<M_*<10^{10}$ M$_{\sun}$ (containing at least 500 star particles) from host haloes with halo masses $M_{\rm 200} > 10^{13}$ M$_{\sun}$, and measure the dark matter fraction, $f_{\rm DM} \equiv M_{\rm DM}(<2 R_{\rm h})/M_{\rm tot}(<2 R_{\rm h})$, within two times of the half-stellar-mass radius, $R_{\rm h}$. Since the EAGLE-highres has $\sim 10$ times better mass resolution than the other two runs, we select its satellite galaxies with $10^8<M_*<10^{10}$ M$_{\sun}$ to have similar effective resolution for stars. Throughout this paper, $M_{200}$ is defined as the mass enclosed in $R_{200}$ within which the average density of a halo is 200 times of the critical density of the universe.

\begin{figure*}
	\includegraphics[width=1.8\columnwidth]{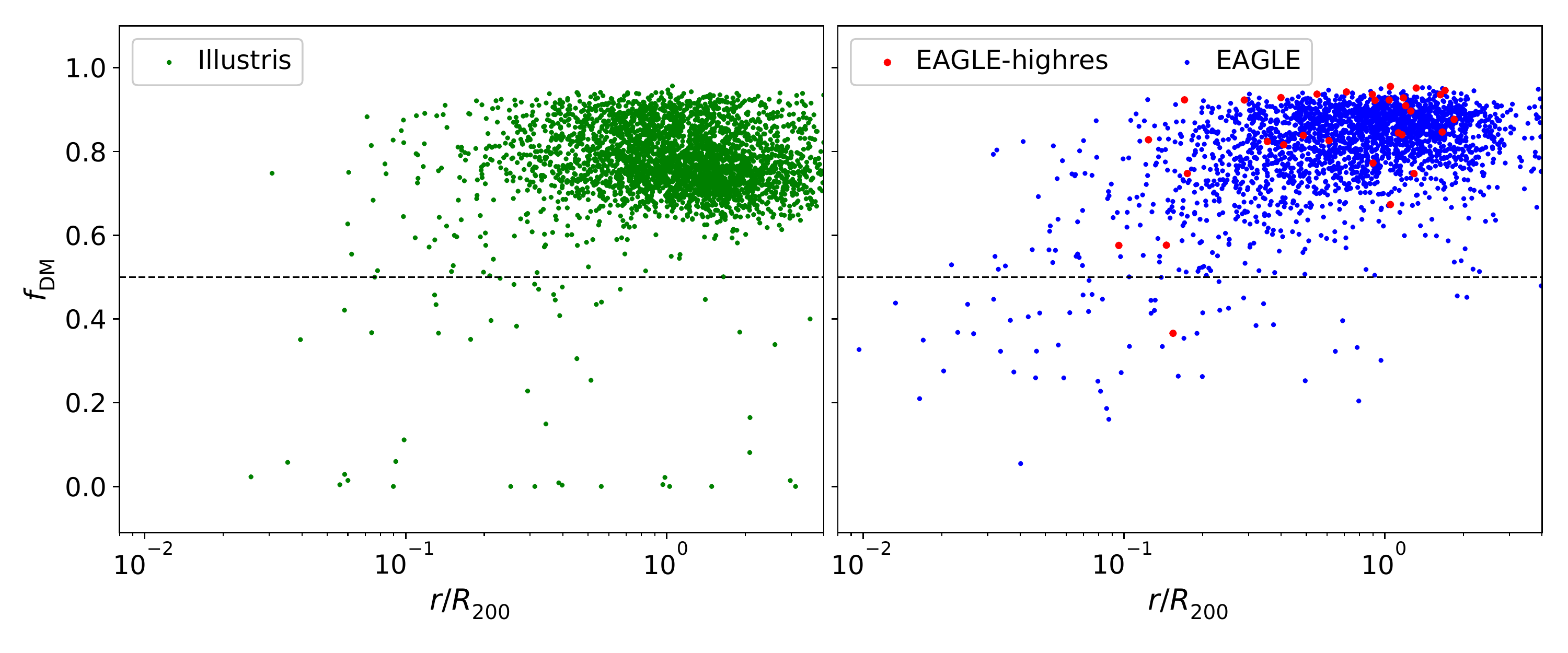}
	\caption{Dark matter fraction $f_{\rm DM}$ as a function of halo-centric distance, $r/R_{200}$, for satellite galaxies in dark matter haloes with $M_{200}>10^{13}$ M$_{\sun}$. Left (Right) panel shows the results of Illustris (EAGLE) satellites. In the right panel, the blue points represent galaxies in EAGLE while the red points denote galaxies in EAGLE-highres. The dashed horizontal lines mark the threshold used to define DMDGs, i.e. $f_{\rm DM}=0.5$. For both Illustris and EAGLE simulations, we select satellite galaxies with stellar masses $10^9<M_*<10^{10}$ M$_{\sun}$, and for EAGLE-highres, we select satellite galaxies with stellar mass $10^8<M_*<10^{10}$ M$_{\sun}$.}
	\label{fig:fig1}
\end{figure*}

In  Fig.~\ref{fig:fig1}, we show the dark matter fraction $f_{\rm DM}$ of our sample galaxies as a function of their halo-centric distances. Results for the Illustris and EAGLE simulations are shown in the left and right panel, respectively. While dark matter dominates the total mass budget of majority satellite galaxies within $2R_{\rm h}$ for all simulations, a few percent of galaxies have $f_{\rm DM}$ below 50\%. 
Here, we define the DMDGs as galaxies with $f_{\rm DM}$ below 50\%.
In Table~\ref{tab:1}, we list the fractions of DMDGs in all satellite galaxies from the Illustris and EAGLE simulations. About 1.5 percent of the Illustris satellite galaxies have dark matter fractions less than 50\% within 2$R_{\rm h}$, while the fraction is about 2.6 percent in the EAGLE simulation.
We note that there are galaxies with $f_{\rm DM}=0$ in Illustris. These galaxies are probably tidal dwarf galaxies formed in tidal tails of interacting galaxies \citep{Ploeckinger2018}, we neglect these galaxies in the rest of this paper.

\begin{table*}
	\centering
	\caption{Fraction of satellite galaxies with $f_{\rm DM}<0.5$ in haloes with $M_{\rm 200}>$ $10^{13}$ M$_{\sun}$. }
	\label{tab1}
	\begin{tabular}{c|c|c|c} 
		\hline
		&Illustris & EAGLE & EAGLE-highres\\
        \hline
		Stellar mass limit (M$_{\sun}$) & $10^9<M_*<10^{10}$ & $10^9<M_*<10^{10}$ & $10^8<M_*<10^{10}$\\
		\hline
		Fraction of DMDGs in satellites  & 1.5\% & ~2.6\% & ~3.2\%\\
        \hline
	\end{tabular}\label{tab:1}
\end{table*}

\begin{figure*}
	\includegraphics[width=2\columnwidth]{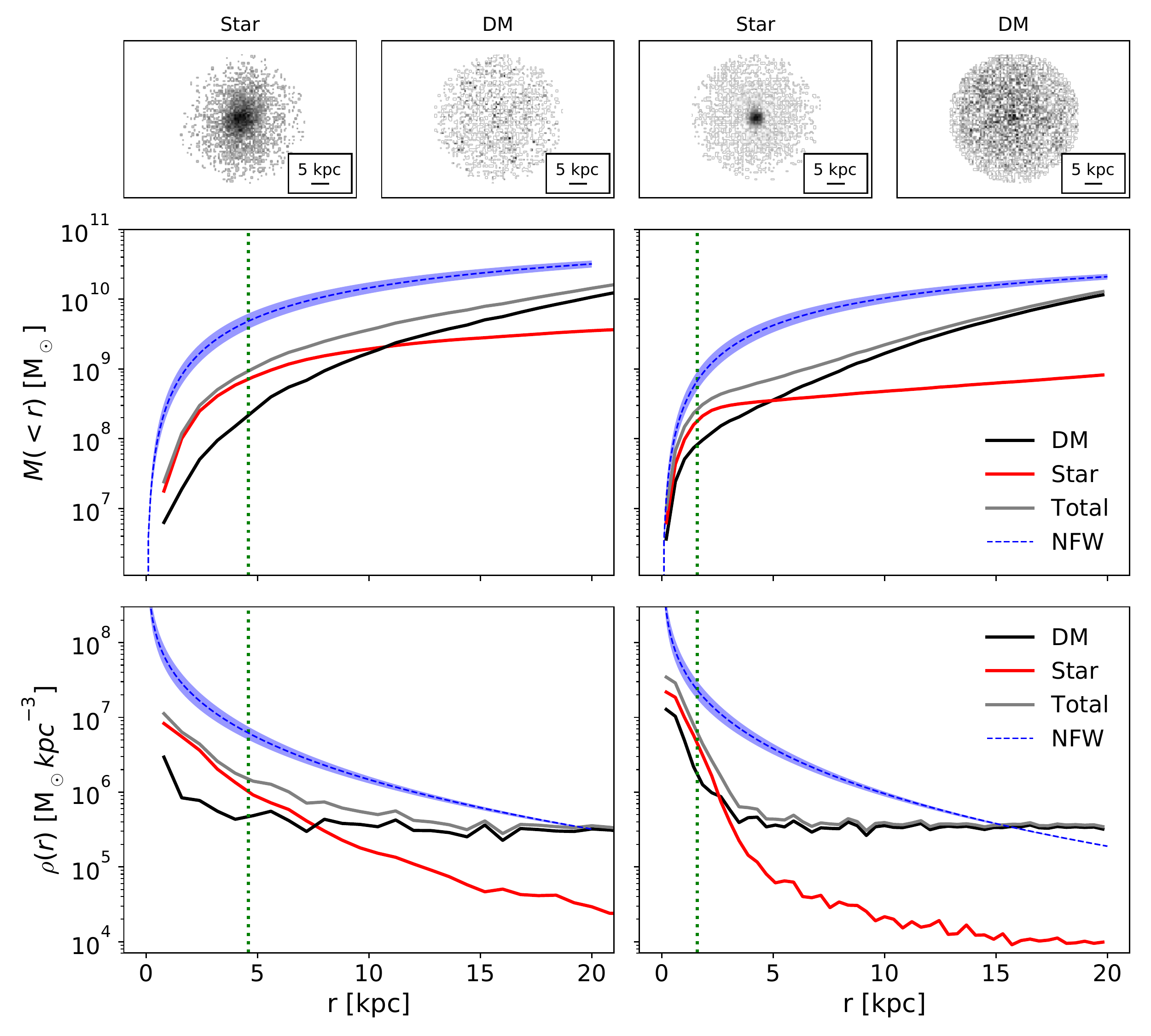}
	\caption{Two representative DMDGs from the Illustris (left) and EAGLE-highres (right) simulations. The top panels show the projected star  and dark matter particles of each galaxy. The middle panels plot the enclosed mass as a function of radius for stars (red), dark matter (black) and total (grey) mass. The bottom panels show density profiles of each component. The green dotted vertical lines indicate the half-stellar-mass radii. The blue dotted lines in the middle and bottom panels show results of NFW halos with masses determined by the mean $M_{*}-M_{200}$ relation in the corresponding simulations ($M_{200}=$ $1.06\times10^{11}$ M$_{\sun}$ (left) and $5.51\times10^{10}$ M$_{\sun}$ (right)). The concentrations are calculated from the concentration--mass relation taken from \citet{Dutton2014} and the blue shade regions represent one sigma of concentration.} 
	\label{fig:fig2}
\end{figure*}

Fig.~\ref{fig:fig2} presents two examples of DMDGs (one from the Illustris and one from the EAGLE-highres). For each galaxy, the projected distributions of star and dark matter particles (upper panels), the cumulative mass profiles (middle panels), and the spherically averaged density profiles of different components (lower panels) are shown. One can clearly see that there are more stars than dark matter within about two times half-stellar-mass radius. 

\section{Formation history of DMDGs}
\label{sec:form}

\begin{figure}
	\includegraphics[width=\columnwidth]{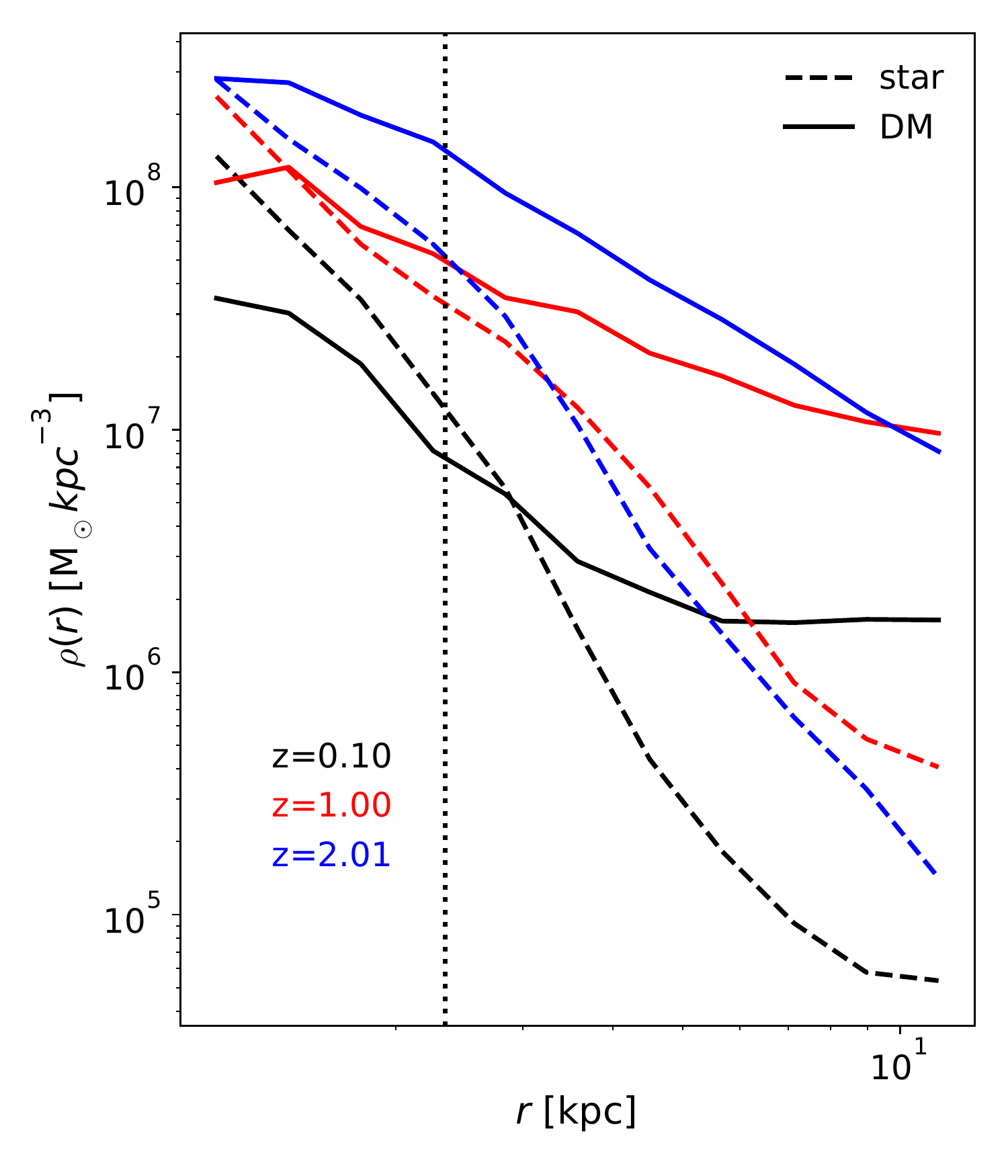}
	\caption{Evolution of density profiles of different components for a DMDG selected from the EAGLE simulation. The dashed and solid lines denote the profiles of stars and dark matter, respectively. The black, red, and bule lines distinguish results for different redshifts as labeled in the figure. The black vertical line indicates $2R_{\rm h}$ of the galaxy at $z=0$}
	\label{fig:profile}
\end{figure}

To understand the origin of these simulated DMDGs, we trace their formation histories back to early time to see their evolutions. In Fig.~\ref{fig:profile}, we present the evolution of density profiles of the dark matter, stars and gas of a representative DMDG selected from the EAGLE at three epochs, $z=0.1, 1.0,$ and $2.0$. Dark matter dominates over stars and gas over all radial range at earlier time, but eventually it becomes subdominant at $z \approx 0$. The shape of dark matter profile does not change much between $z=1.0$ and $z=0.1$, while its amplitude drops by a factor of ten. This suggests the mass loss of dark matter occurred at all scales, consistent with \citet{Frings2017}, which used idealized simulations and found the whole dark matter profile of satellite decreases due to tidal forces after infall. In the same redshift range, the amplitude of stellar profile only decreases by 10-50\%. Apparently the mass loss of dark matter is more rapid than that of stars.

\begin{figure}
	\includegraphics[width=\columnwidth]{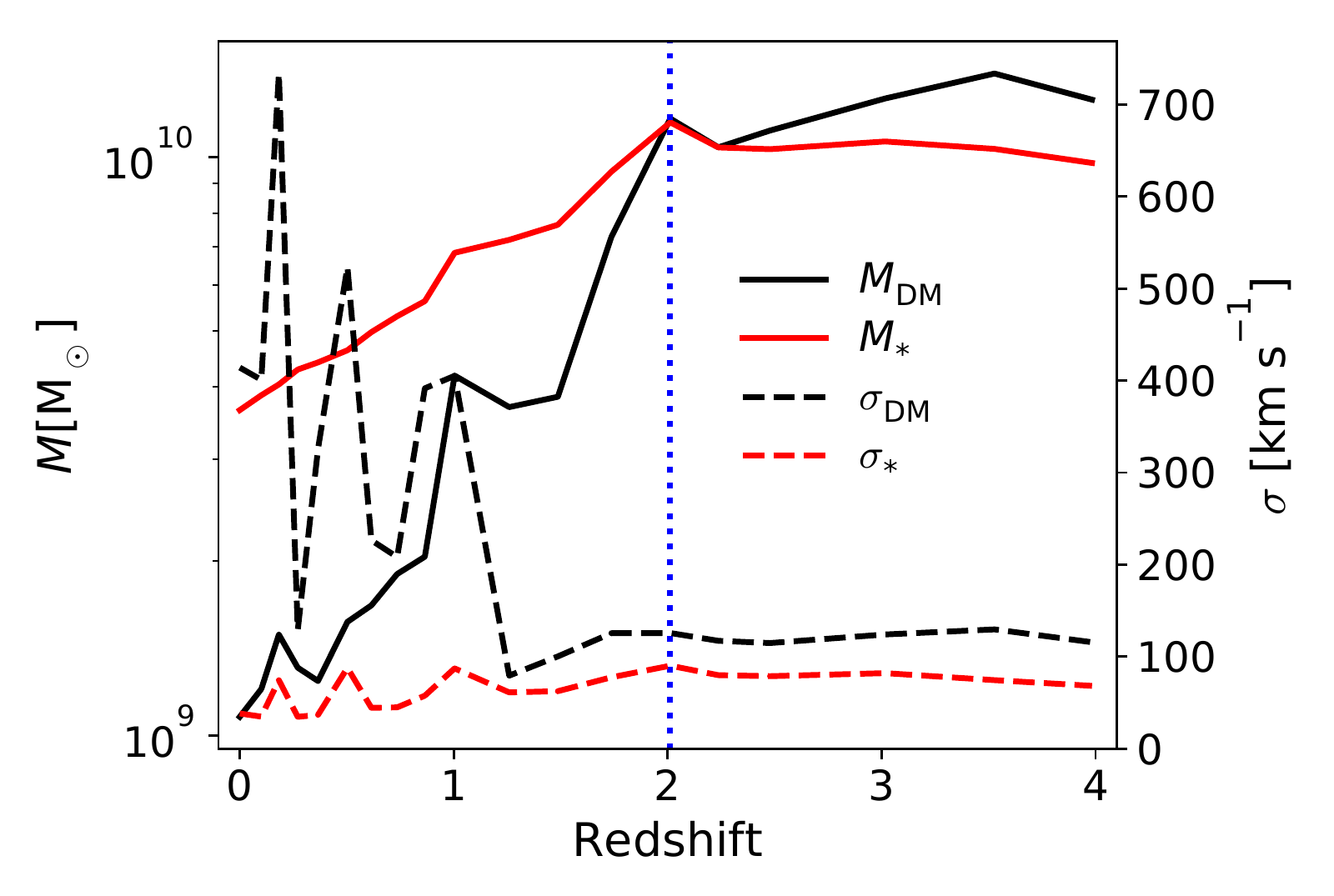}
	\caption{Mass (left axis) and velocity dispersion ($\sigma$) (right axis) of dark matter and star components as a function of redshift for the DMDG shown in Fig.~\ref{fig:profile}. The mass and dispersion at each redshift are calculated within 2.34 physical kpc ($2R_{\rm h}$ of the galaxy at $z=0$). The black and red solid lines represent enclosed mass results for the dark matter and stars, respectively. The black and red dashed lines denote the velocity dispersion results for the dark matter and stars, respectively. The blue vertical line indicates the infall time of the satellite galaxy.}
	\label{fig:growth}
\end{figure}

This can be further shown in Fig.~\ref{fig:growth} where we plot the mass growth history (left axis) of the same galaxy. Here the masses of galaxies are defined as the enclosed mass within fixed 2.34 physical kpc ($2R_{\rm h}$ of the galaxy at $z=0$). The blue vertical line indicates the infall time of the satellite galaxy, which is defined as the redshift when the galaxy enters into $R_{200}$ of its parent halo. Clearly, both the enclosed dark matter and stars within the range decrease rapidly after the infall redshift $z_{\rm infall} \sim 2$, yet dark matter loses mass at a much higher rate than the stellar component. In the end, for both components, only a few percent of their initial masses remain. 

What cause the galaxy to lose dark matter more rapidly than stars? One possibility is that dark matter and star particles have different energy distributions. To verify this, in the same figure we plot the velocity dispersion of dark matter particles and star particles within the same $r$ range. The velocity dispersion of dark component is about $30\%$ higher than that of stars at early time, and the difference increases significantly at later time, which may be induced by the tidal shocking \citep[e.g.][]{Gnedin1997,Gnedin1999a,Gnedin1999b,Kazantzidis2011,Prieto2008}. This supports the expectation that dark matter particles are dynamically ``hotter'' than stars and so are prone to be tidally disrupted. We demonstrate it further below.

\begin{figure}
	\includegraphics[width=\columnwidth]{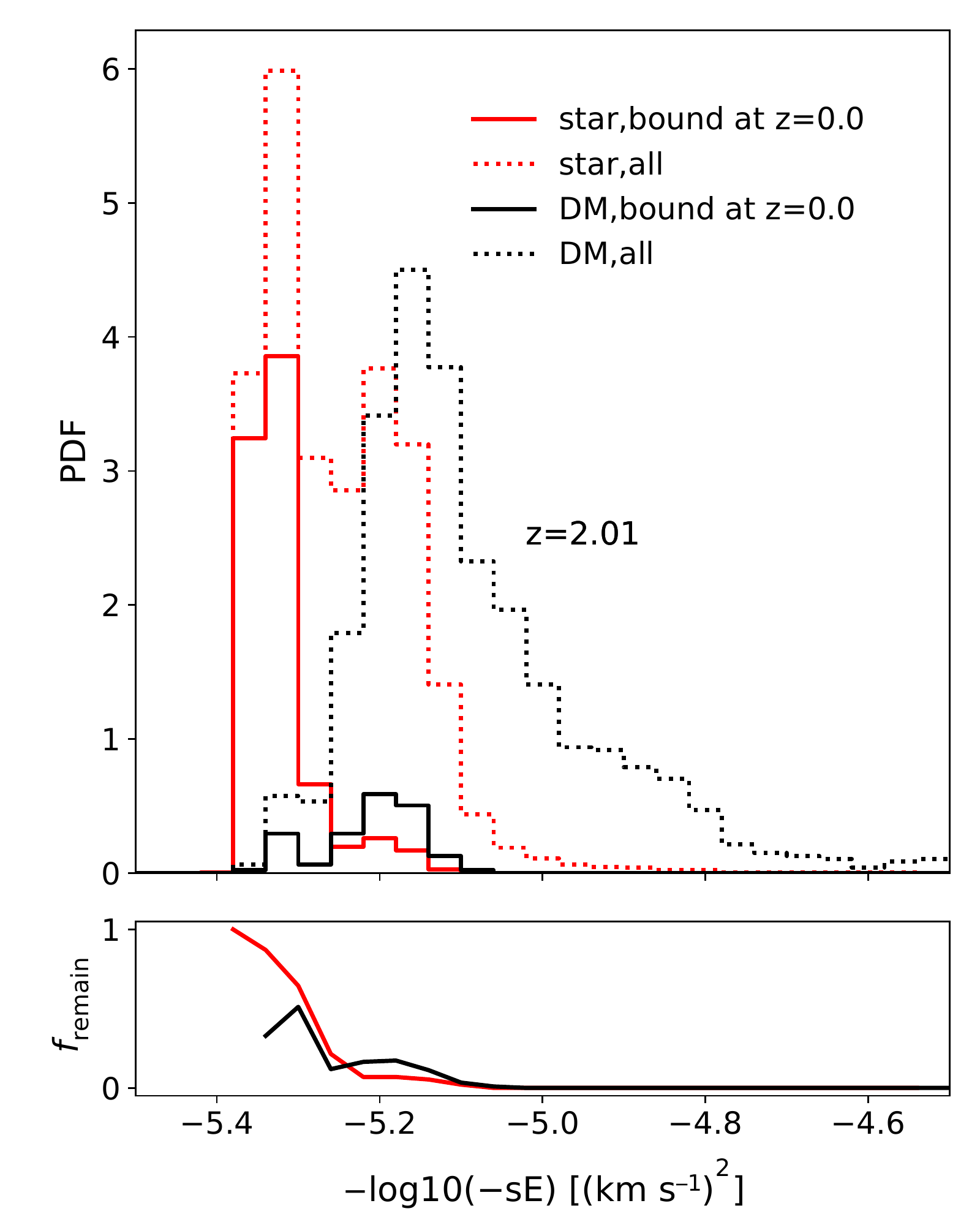}
	\caption{Top panel: the dotted histograms show the mass weighted and normalized specific binding energy distributions of dark matter (black) and stars particles (red) within 2.34 physical kpc at $z=2.01$ for the DMDG shown in Fig.~\ref{fig:profile}. The solid histograms show results for the dark matter and star particles remaining bound at $z=0$. The lower panel shows the fraction of mass that still remains bound at $z=0$ in each $sE$ bin.}
	\label{fig:bind_infall}
\end{figure}

We select all the star/dark matter particles within 2.34 physical kpc of the galaxy at $z=2.01$ ($z_{\rm infall}$) at which the galaxy enters into the virial radius of its parent halo. Then we plot their specific binding energy distributions in Fig.~\ref{fig:bind_infall} as dashed lines. Different colors distinguish the distributions of different components. Here, the specific binding energy $sE$, i.e. energy per mass, of particle $i$ is calculated by 
\begin{equation}
 sE_i=-\sum_{j\neq i} \frac{G m_{j}}{r_{i,j}}+\frac{1}{2}v_{i}^2,   
\end{equation}
 where $m_j$ denotes the mass of particle $j$, $r_{i,j}$ is the distance from particle $i$ to particle $j$, and $v_{i}$ is the velocity of particle $i$ with respect to the center-of-mass velocity. The summation here accounts for all particles/cells belonging to this SUBFIND galaxy except for particle $i$.
Clearly the dark matter and star particles have different binding energy distributions, and hence these two components should suffer different tidal impacts. 

We also plot the specific energy distribution of the particles which still remain to be bound at $z=0$ in the same plot with solid lines. Comparing to the results at $z=2$,  only the particles with low binding energy remain at the present day. In the lower panel of this figure, we show the fraction of mass that still remains bound at $z=0$ as a function of $sE$, clearly particles with higher binding energy tend to be tidally disrupted.

The tidal origin of DMDGs has been discussed in literature \citep{Ogiya2018, Chang2013, Yu2018}. \citet{Ogiya2018} argued that the more rapid mass loss of dark matter could be explained if a galaxy has a cored mass distribution before infall. We find that both EAGLE and Illustris simulations have cuspy dark matter distribution in dwarf galaxies. Recent studies demonstrate that this may be due to 
the low gas density threshold for star formation  \citep[e.g.][]{Benitez-Llambay2018,Bose2018,Dutton2019}. Therefore cored profiles are not a necessary condition to form DMDGs.

Up to now, we mainly focus on the discussion of one typical DMDG example. In Appendix \ref{ap:a}, we show 11 additional examples, five from Illustris, five from EAGLE, and one from EAGLE-highres. In addition, we also show the stacked results of Illustris and EAGLE in Appendix \ref{ap:a}. Our conclusions do not change when considering all DMDGs in the sample.

\begin{figure}
	\includegraphics[width=\columnwidth]{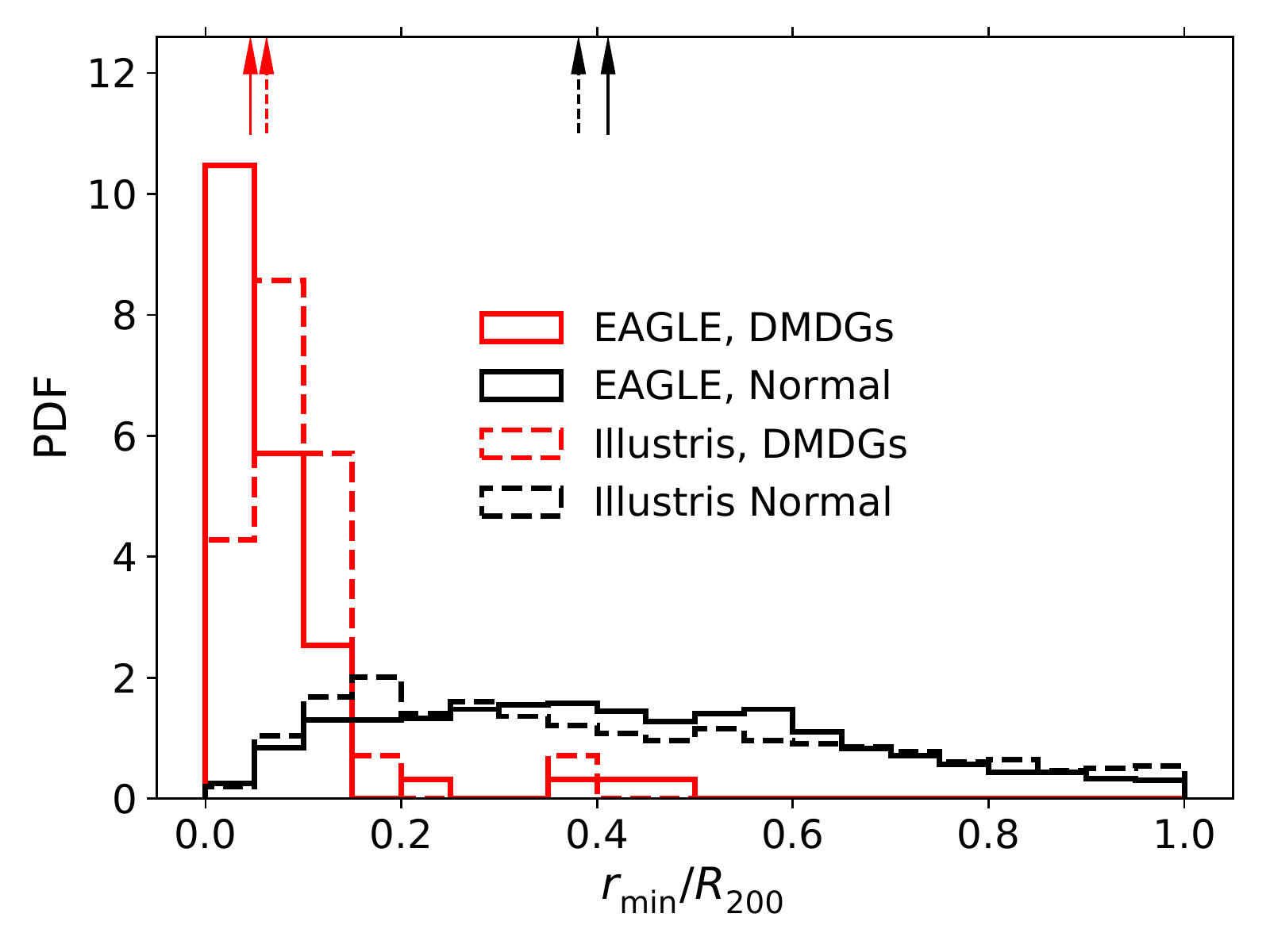}
	\caption{Distributions of $r_{\rm min}/R_{200}$ for DMDGs (red) and normal (black) satellite galaxies in the EAGLE simulation. The arrows denote the corresponding median values of $r_{\rm min}/R_{200}$.}
	\label{fig:peri}
\end{figure}

The galaxy discussed above apparently suffers very strong tidal interactions with its host. Do all DMDGs once experience such strong tidal interactions? In Fig.~\ref{fig:peri}, we plot the distribution of $r_{\rm min}$ for the DMDGs in the EAGLE and Illustris simulations, and compare it with that for the normal satellite galaxies. Here $r_{\rm min}$ is an approximation of the pericenter distance, which is defined as the minimal distance of the galaxy from its parent halo in its history. Generally, a smaller $r_{\rm min}$ corresponds to stronger tidal interactions. We find that DMDGs have a very different $r_{\rm min}$ distribution from normal satellite galaxies, i.e. in the EAGLE (Illustris) simulation, DMDGs' $r_{\rm min}$ tend to distribute in the very inner region of their host haloes with a median value of $r_{\rm min}/R_{200} = 0.046$ (0.062), which is significantly smaller than that of normal satellite galaxies, $r_{\rm min}/R_{200} = 0.411$ (0.381).

Although the pericenter distance is usually a good estimator of the degree of tidal interaction, the tidal stripping strength also depends on other orbital parameters, such as the orbital energy at infall, the time spent in host halo and the eccentricity of orbit, and the properties of satellite, such as its stellar mass, and its compactness \citep[see e.g.][]{Frings2017,Fattahi2018,Buck2019,Li2019}. 

\begin{figure}
	\includegraphics[width=\columnwidth]{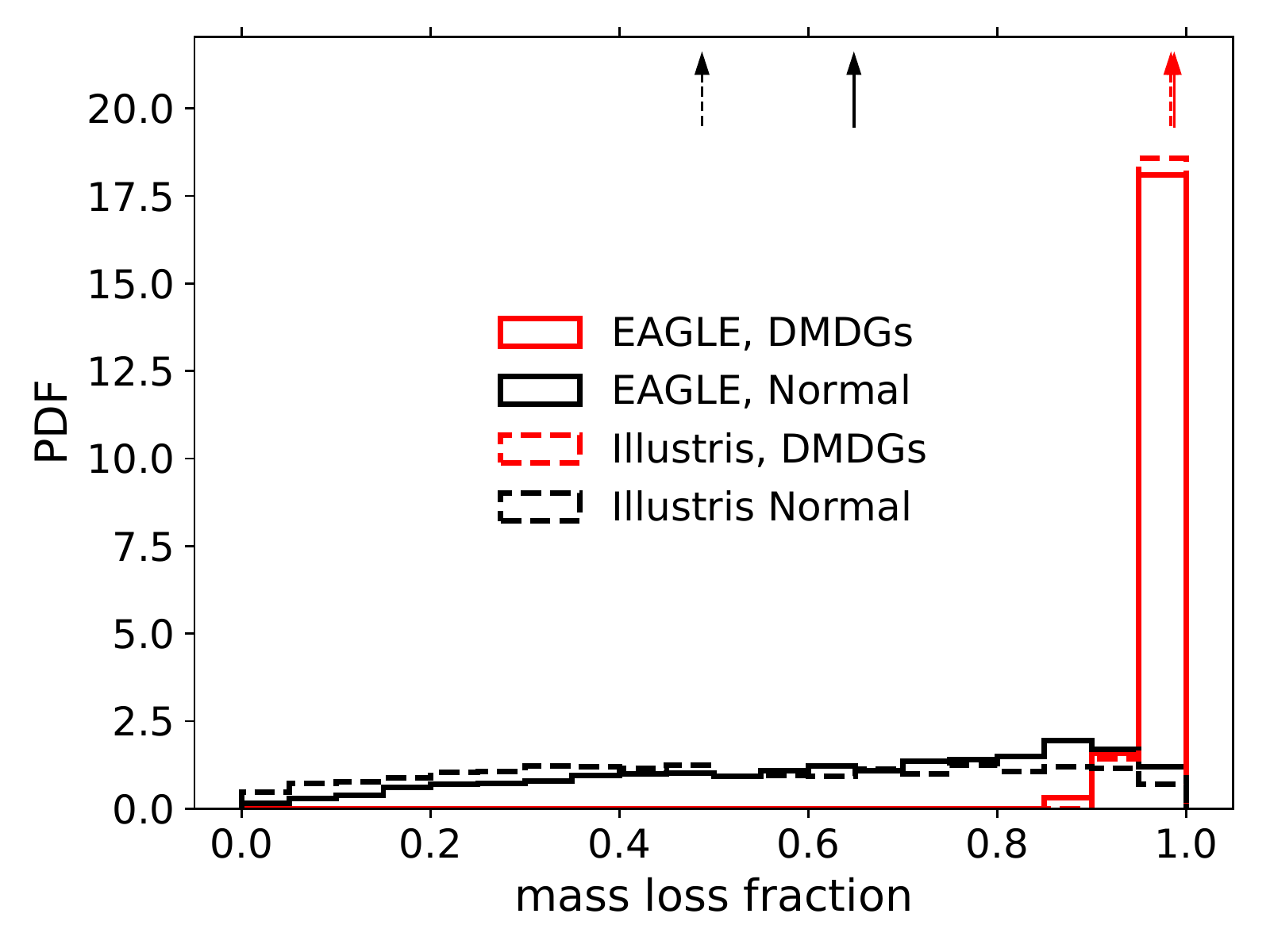}
	\caption{Distributions of the mass loss fractions for DMDGs (red) and normal satellite galaxies (black) in the EAGLE simulation. The galaxy sample is the same as that in Fig.~\ref{fig:peri}. The arrows denote the corresponding median mass loss fractions.}
	\label{fig:massloss}
\end{figure}

Instead of exploring each of these parameters. We directly compare in  Fig.~\ref{fig:massloss} the mass loss fractions of the DMDGs and the normal satellite galaxies. The mass loss fraction is calculated as the total SUBFIND mass loss of the satellites since their infall. It is easy to find that DMDGs all suffer significant mass loss. The median mass loss fraction for DMDGs is 98.7\% (98.4\%), while it is 64.8\% (48.7\%) for normal satellite galaxies in EAGLE (Illustris) simulation. All these results strongly suggest that DMDGs are highly tidal disrupted galaxies.

\section{Conclusions}
\label{sec:sum}
We use two sets of state-of-the-art hydrodynamical cosmological simulations to investigate whether or not DMDGs are allowed in current galaxy formation models. Our results can be summarized as follows.

For satellite galaxies in galaxy groups/clusters, 1.5\% of galaxies have dark matter fractions below 50\% within two times half-stellar-mass radius in the Illustris simulation, while it is slightly more abundant in the EAGLE, which is 2.6\%.

We trace the formation histories of DMDGs back to earlier time, and find these DMDGs are not originally dark matter deficient but become so at later time due to strong tidal interactions with central galaxies of their hosts. During the interactions, because the binding energy of dark matter in these DMDGs is significantly higher than stars, the mass loss of dark matter is much more rapid than stars, leading to be dark matter deficient at the present day. Comparing to normal satellite galaxies in group/clusters, on average, the median value of $r_{\textrm{min}}/R_{200}$ of DMDGs is two times closer to their central galaxies; DMDGs also lose a much larger fraction of their original masses than normal satellite galaxies. These results suggest that DMDGs are highly disrupted systems, and if DMDGs were confirmed in observations, they are expected in current galaxy formation models. 

\section*{Acknowledgements}
We thank the anonymous referee for the valuable comments. We acknowledge supports from the National Key Program for Science and Technology Research and Development of China (2017YFB0203300, 2015CB857005), the National Natural Science Foundation of China (Nos. 11425312,11503032, 11773032, 11390372, 11873051, 118513, 11573033, and 11622325), the NAOC Nebula Talents Program, and the Newton Advanced Fellowship.

\bibliographystyle{mnras}
\bibliography{mnras} %


\appendix

\section{Addition examples} \label{ap:a}
\begin{figure*}
	\includegraphics[width=2\columnwidth]{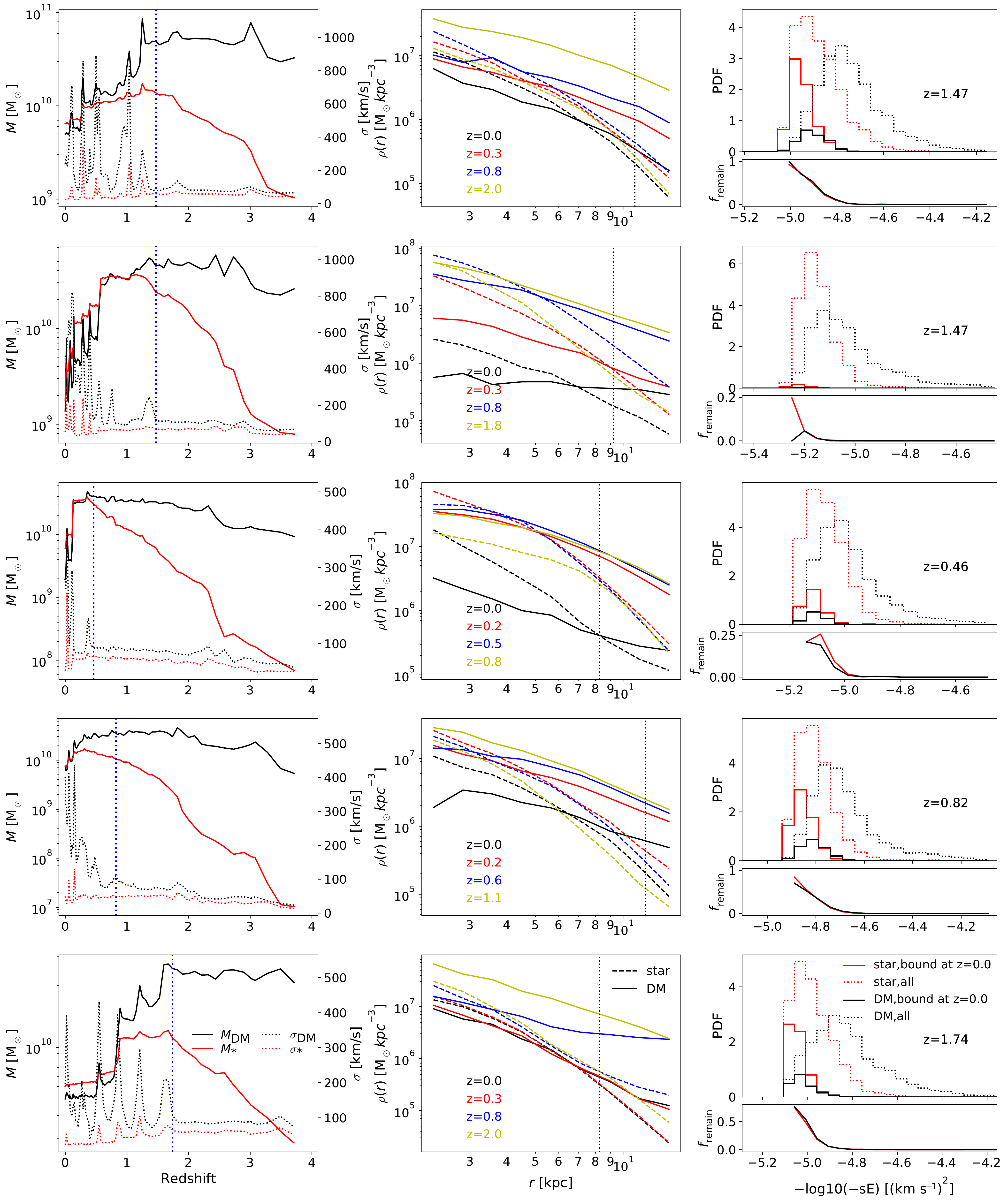}
	\caption{Evolution of DMDGs examples selected from Illustris. Each row denotes one DMDG. The columns from left to right are similar to Fig.~\ref{fig:growth}, \ref{fig:profile} and \ref{fig:bind_infall}, respectively.}
	\label{fig:A1}
\end{figure*}

\begin{figure*}
	\includegraphics[width=2\columnwidth]{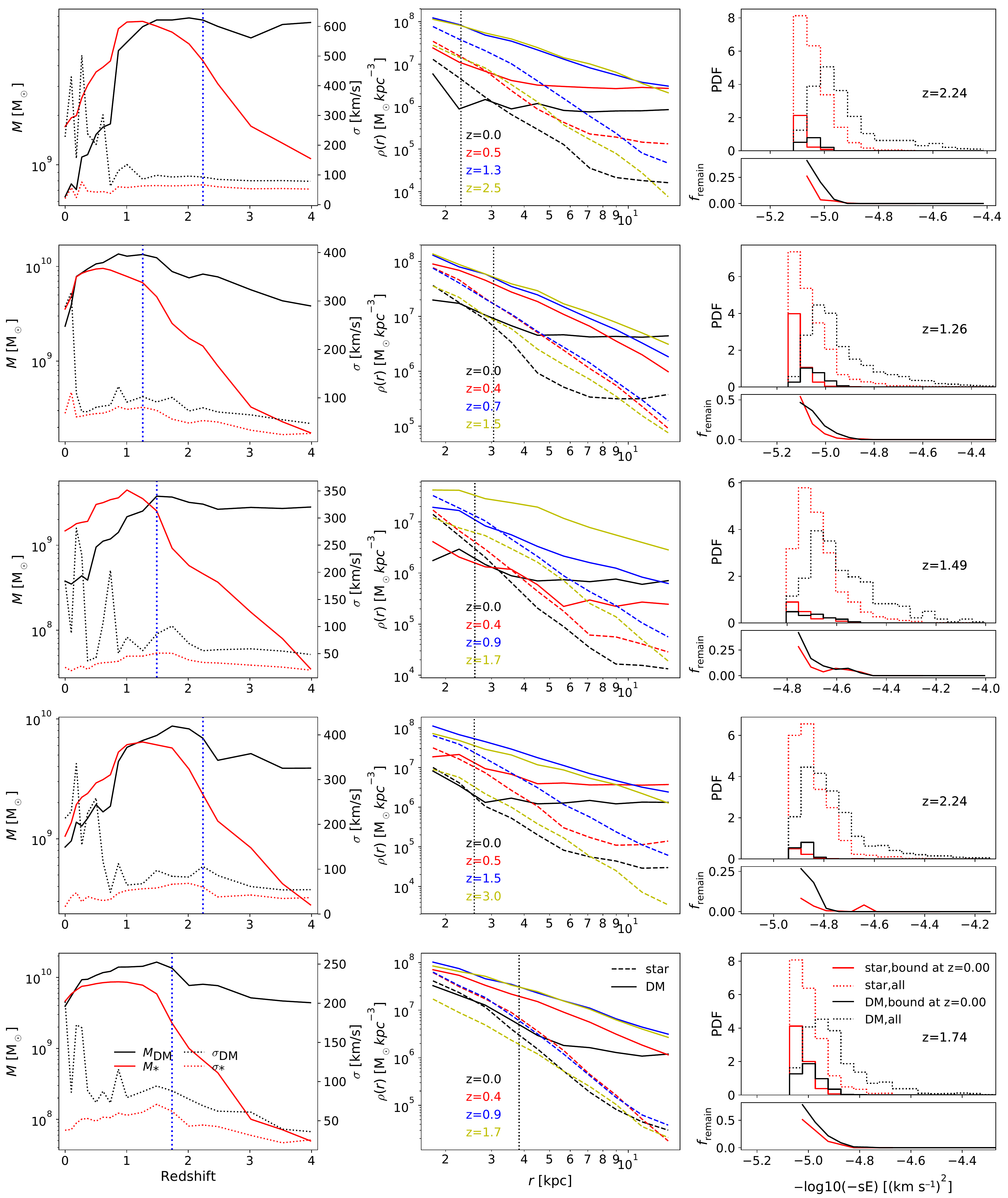}
	\caption{Similar to Fig~\ref{fig:A1}, but selected from EAGLE.}
	\label{fig:A2}
\end{figure*}

\begin{figure*}
	\includegraphics[width=2\columnwidth]{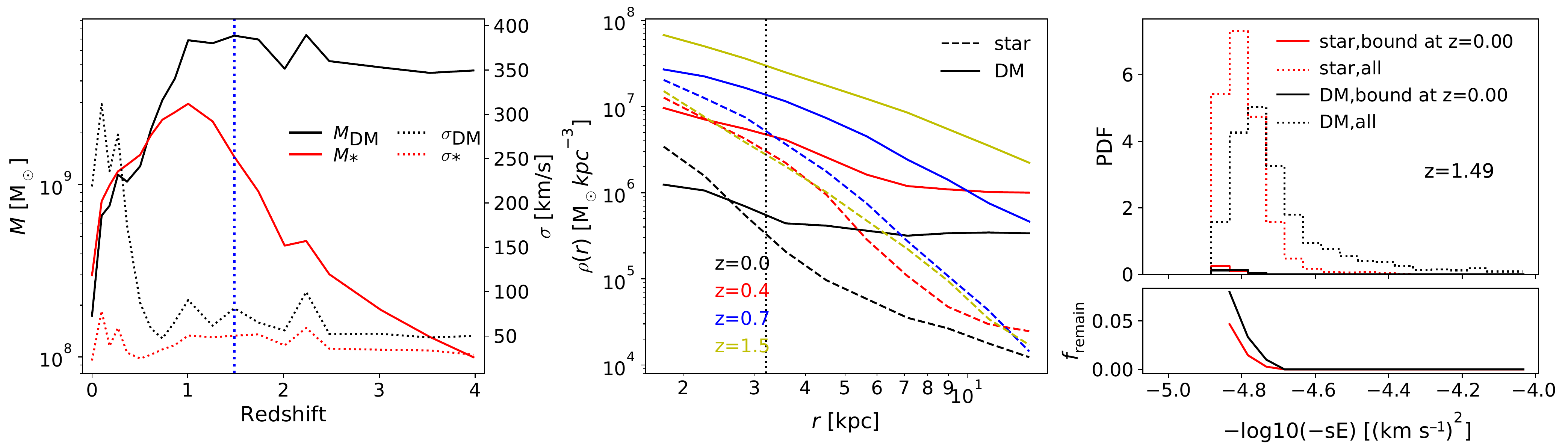}
	\caption{Similar to Fig~\ref{fig:A1}, but selected from EAGLE-highres.}
	\label{fig:A3}
\end{figure*}

\begin{figure*}
	\includegraphics[width=2\columnwidth]{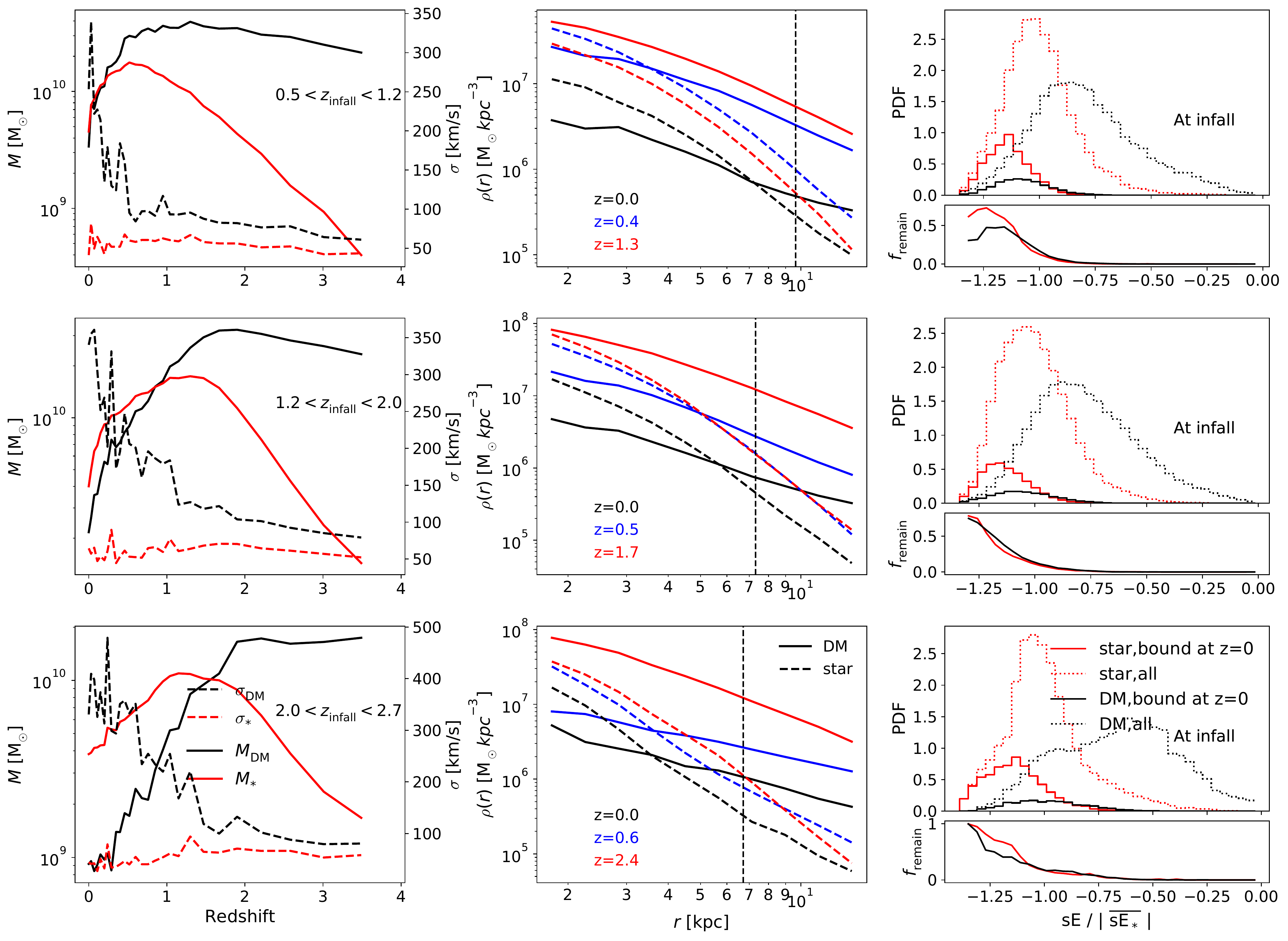}
	\caption{Stacked results for DMDGs within different $z_{\rm infall}$ bins for Illustris. Each row represents the stacked results of DMDGs in the same $z_{\rm infall}$ bin. Left columns are similar to Fig.~\ref{fig:growth}, but the lines show the mean value of the DMDGs. Middle columns are similar to Fig.~\ref{fig:profile}, but the lines show the mean density profiles. Right columns are similar to Fig.~\ref{fig:bind_infall}, but histograms show the stacked distribution of $sE$ which is normalized by the mean $sE$ of the star particles of each galaxy before stacking.}
	\label{fig:A4}
\end{figure*}

\begin{figure*}
	\includegraphics[width=2\columnwidth]{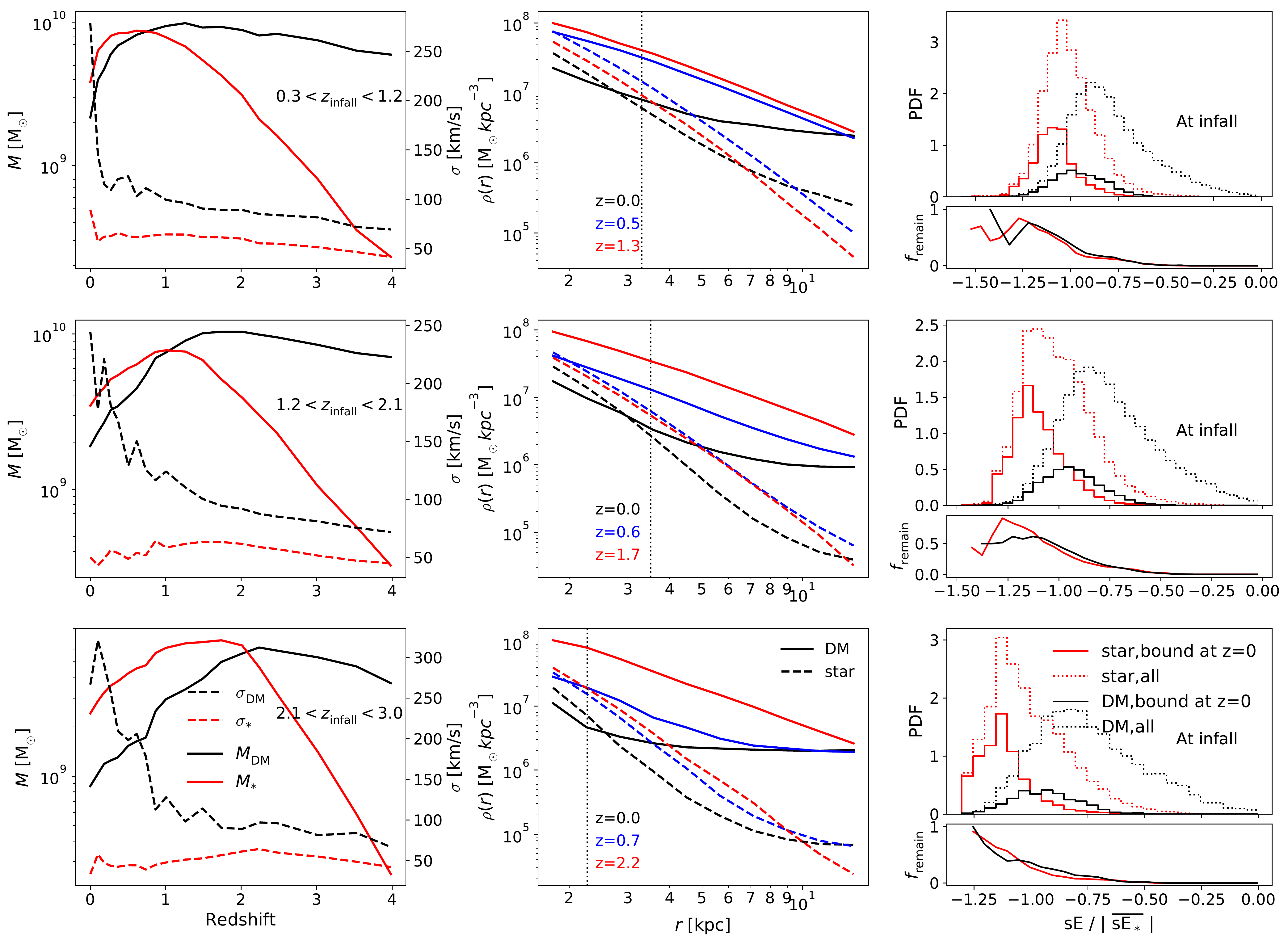}
	\caption{Similar to Fig.~\ref{fig:A4}, but for EAGLE.}
	\label{fig:A5}
\end{figure*}

Here we show 11 additional examples, five from Illustris (Fig.~\ref{fig:A1}), five from EAGLE (Fig.~\ref{fig:A2}), and one from EAGEL-highres (Fig.~\ref{fig:A3}). Each row denotes one DMDG. The columns from left to right are similar to Figs~\ref{fig:growth}, \ref{fig:profile} and \ref{fig:bind_infall}, respectively. We also show the stacked results of Illustris and EAGLE in Fig.~\ref{fig:A4} and Fig.~\ref{fig:A5}, respectively.

\bsp	
\label{lastpage}
\end{document}